\documentstyle[twoside,fleqn,npb,epsfig]{article}
\newcommand{\AmS}{{\protect\the\textfont2
  A\kern-.1667em\lower.5ex\hbox{M}\kern-.125emS}}
\newcommand{\SMALLX}{SMALLX}
\newcommand{\CCFM}{CCFM}
\newcommand{\BFKL}{BFKL}
\newcommand{\LDC}{LDC}
\newcommand{\alphasb}{\bar{\alpha}_s}

\title{CCFM prediction for $F_2$ and forward jets at HERA}

\author{H. Jung\address{Department of Physics \\ 
Lund University\\
221 00 Lund, Sweden\\ 
E-mail: jung@mail.desy.de}%
}

\begin{document}

\begin{abstract}
Predictions of the CCFM evolution equation for $F_2$ and forward jets at HERA
energies are obtained from a modified version of the Monte Carlo program SMALLX. 
The treatment of the non-Sudakov form factor $\Delta_{ns}$ is discussed as well
as the effect of the so called ``consistency constraint". For the first time a
good description of $F_2$ and the forward jet data is obtained from the CCFM
equation.

\end{abstract}

\maketitle

\section{Introduction}
The parton evolution at small values of $x$ is believed to be best described by
the CCFM evolution equation \cite{\CCFM}, which for $x \to 0$ is equivalent to
the BFKL evolution equation \cite{\BFKL} and for large $x$ reproduce the
standard DGLAP equations. The CCFM evolution equation takes coherence effects
of the radiated gluons into account via angular ordering. On the basis of this
evolution equation, the Monte Carlo program SMALLX \cite{\SMALLX} has been
developed already in 1992. 
 In 1997 the Linked Dipole Chain \cite{\LDC} was 
developed as a reformulation of the original CCFM equation.
Predictions of the CCFM equation for hadronic final state properties were
studied in~\cite{Salam}
paying special attention to non - leading effects. 
All approaches~\cite{\LDC,Salam} found a good description of $F_2$ but
failed completely to describe the forward jet data of HERA experiments, which
are believed to be a signature for new small $x$ parton dynamics.
\par
In the following I  discuss  the
treatment of the non-Sudakov form factor $\Delta_{ns}$ as well 
as the effects of
the so-called ``consistency constraint",
 which was found to be necessary to include
non - leading contributions to the BFKL equation \cite{Martin}. I show that
a good description of $F_2$ and the forward jet data can be achieved.

\section{Implementation of CCFM in SMALLX }
The implementation of the CCFM~\cite{\CCFM} parton evolution in the forward
evolution Monte Carlo program SMALLX is described in detail in~\cite{\SMALLX}. 
Here I only concentrate on the basic ideas and discuss the treatment of the
non-Sudakov form factor.
\par
The initial state gluon cascade is generated 
in a forward evolution approach from a starting distribution of the
$k_t$ unintegrated gluon distribution according to:
\begin{equation}
x G_0(x,k_t^2) = N \cdot (1-x)^4 \cdot \exp{\left(-k_t^2/k_0^2\right)}
\end{equation}
where $N$ is a normalization constant and $k_0^2=1$~GeV$^2$. 
Gluons then can
branch into a virtual ($t$-channel) gluon $k_{i+1}$ and a final gluon $q_{i+1}$
according to the CCFM splitting function \cite{\CCFM}:
\begin{eqnarray}
d{P}_i & =& \tilde{P}_g^i(z_i,q^2_{ti},k^2_{ti}) \cdot \Delta_s d z_i
             \frac{d^2 q_{ti} '}{\pi q'^{2}_{ti}} \nonumber \\
	 &  &	 \cdot \Theta(q_{ti}'-z_iq_{t\;i-1}')
		 \cdot \Theta(1-z_i-\epsilon_i)
\end{eqnarray}		 
with $q'_{ti}=q_{ti}/(1-z_i)$ being the rescaled transverse momentum,
$z_i=x_i/x_{i-1}$, $\epsilon_i=Q_0/q'_{ti}$ being a collinear cutoff to avoid
the $1/(1-z)$ singularity and
$\Delta_s$ being the Sudakov form factor:
\begin{equation}
\Delta_s(q'_{ti},z_iq'_{t\;i-1}) =\exp{\left(
 -\int \frac{d^2 q'_t}{q'^2_t} \int dz \frac{\alphasb}{1-z}
 \right)}
 \end{equation}
which at an inclusive level cancels 
against the $1/(1-z)$ collinear singularity and
 is used to generate $q'_{ti}$.
The gluon splitting function $\tilde{P}_g^i$ is given by:
\begin{equation}
\tilde{P}_g^i= \frac{\alphasb(q^2_{ti})}{1-z_i} + 
\frac{\alphasb(k^2_{ti})}{z_i} \Delta_{ns}(z_i,q^2_{ti},k^2_{ti})
\end{equation}
with the non-Sudakov form factor $\Delta_{ns}$ being defined as:
\begin{eqnarray}
\log\Delta_{ns} & = & -\alphasb(k^2_{ti})
                  \int \frac{dz'}{z'} 
			\int \frac{d q^2}{q^2} \nonumber \\ 
              & & \cdot \Theta(k_{ti}-q)\Theta(q-z'q_{ti})			
\label{ns}
\end{eqnarray}
which gives for the region $k^2_{ti} > z_i q^2_{ti}$:
\begin{equation}
\log\Delta_{ns} = -\alphasb(k^2_{ti})
\log\left(\frac{1}{z_i}\right)\log\left(\frac{k^2_{ti}}{z_i q^2_{ti}}\right)
\label{ns_simple}
\end{equation}
 The constraint 
$k^2_{ti} > z_i q^2_{ti}$ is often referred to
 as the ``consistency constraint"~\cite{Martin}.
However the upper limit of the $z'$ integral 
is constraint by the $\Theta$ functions in eq.(\ref{ns}) 
by\footnote{I am grateful to J. Kwiecinski for the explanation of this
constraints}:
$z_i \leq z^{\prime} \leq \mbox{min}(1,k_{ti}/q_{ti}) $, which results in the
following form of the non-Sudakov form factor \cite{Martin_Sutton}:
\begin{equation}
\log\Delta_{ns} = -\alphasb(k^2_{ti})
\log\left(\frac{z_0}{z_i}\right)
\log\left(\frac{k^2_{ti}}{z_0z_i q^2_{ti}}\right)
\label{ns_new}
\end{equation} 
where
$$z_0 = \left\{ \begin{array}{ll}
              1             & \mbox{if  } k_{ti}/q_{ti} > 1 \\
		  k_{ti}/q_{ti} & \mbox{if  } z < k_{ti}/q_{ti} \leq 1 \\
		  z             & \mbox{if  } k_{ti}/q_{ti} \leq z  
		  \end{array} \right. $$
giving no supression
 in the region $k_{ti}/q_{ti} \leq z$ we have $\Delta_{ns}=1$.
The Monte Carlo program SMALLX~\cite{\SMALLX} has been modified to include the
non-Sudakov form factor according to eq.(\ref{ns_new}) and the scale in
$\alpha_s$ was changed to $k_{t}^2$ in the cascade and 
in the matrix element. To avoid problems at small $k_t^2$,
 $\alpha_s (k_t^2)$ is restricted to
 $\alpha_s (k_t^2) \leq 0.6$. 
\par
The ``consistency constraint"
was introduced to
account for next-to-leading effects in the BFKL equation, and which 
was found 
\cite{Martin} to simulate about 70\% of the full
next-to-leading corrections to the BFKL equation. Since 
in LO BFKL the true kinematics of the branchings are neglected, they 
can be interpreted
as next-to-leading effects, and this constraint is often also called
``kinematic constraint". In the CCFM equation energy and momentum conservation
is already included at LO, and it is not clear, whether the arguments coming
from BFKL also apply to CCFM. In the following the effects of the $1/(1-z)$
terms and the ``consistency constraint" are studied in more detail.

\section{Predictions for $F_2$ and forward jets at HERA}

\begin{figure}[htb]
\vskip -15mm 
\epsfig{figure=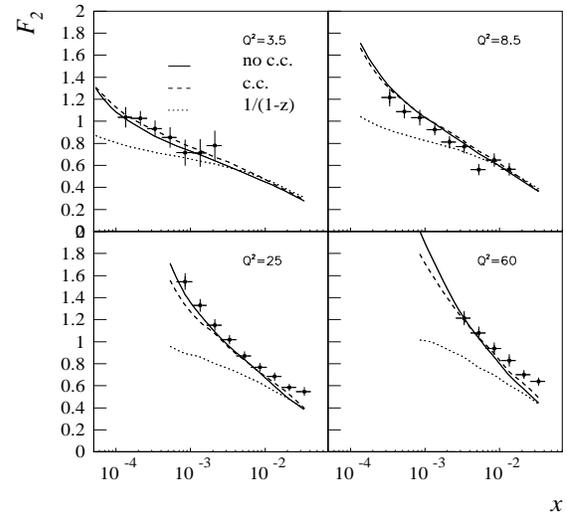,width=8.5cm,height=8cm}
\vskip -10mm 
\caption{
The structure function $F_2(x,Q^2)$ compared to 
H1 data~\protect\cite{H1_F2_1996}.
 The solid (dashed) line is 
the prediction of the Monte Carlo without (with)
applying the ``consistency constraint" (c.c.)
and the dotted line includes only the $1/(1-z)$ term.
  }\label{f2_1}
\end{figure}

\begin{figure}[htb]
\vskip -12mm 
\epsfig{figure=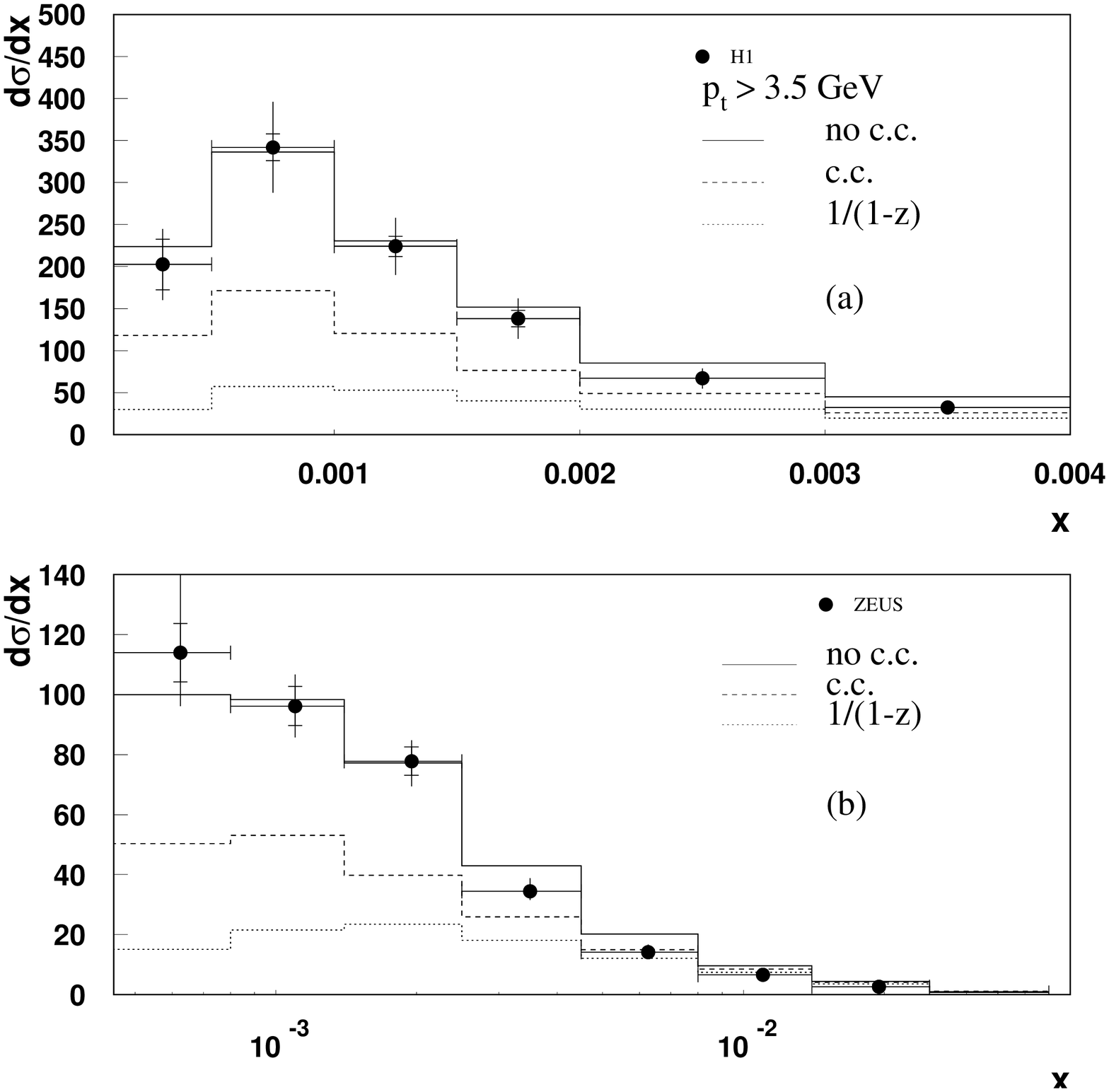,
width=8cm,height=10cm}
\vskip -8mm 
\caption{
 The cross section for forward jet production as a function of $x$, 
 compared to H1 data~\protect\cite{H1_fjets_data}
($a.$) and 
compared to ZEUS data~\protect\cite{ZEUS_fjets_data} ($b.$).
 The solid (dashed) line is 
the prediction of the Monte Carlo without (with)
applying the ``consistency constraint" (c.c.)
and the dotted line includes only the $1/(1-z)$ term.
  }\label{fwd_jets_1}
\end{figure}

With the modifications on the treatment of the non-Sudakov form factors
described above, the predictions of the CCFM evolution equation, 
as implemented
in the program SMALLX~\cite{\SMALLX}, for the structure function 
$F_2 (x,Q^2)$ are
shown in Fig.~\ref{f2_1} without applying any additional ``consistency
constraint". 
Here the following parameters were used: $Q_0=1.1$~GeV,
$\Lambda_{QCD} = 0.2 $~GeV, $N=0.4$ and 
the masses for light (charm) quarks were set to
$m_q=0.25 (1.5)$~GeV. The scale of $\alpha_s$ in the 
off-shell matrix element was set
to $k_t^2$. With this parameter settings are very good description
 of $F_2$ over
the range $0.5 \cdot 10^{-5} < x < 0.05$ and $3.5 < Q^2 < 90$~GeV$^2$ 
is obtained. 
In Fig.~\ref{fwd_jets_1} the prediction for the forward jets using the same
parameter setting is shown. The data are nicely described.
The effect of the $1/(1-z)$ term in the splitting function is shown in 
Figs.~\ref{f2_1} and \ref{fwd_jets_1} separately with the dashed line. 
It is obvious that these terms are
important for a reasonable description of $F_2$ and the forward jet data.
\par
Including the ``consistency constraint" the $x$ dependence of the cross section changes, but
a similarly good description of $F_2$ is obtained by changing $Q_0$ to $Q_0=0.85$ GeV. However the
forward jet cross section becomes smaller,
 as shown with the dotted line
 in Figs.~\ref{f2_1} and \ref{fwd_jets_1}. 
 It is interesting to note, that
this prediction is very similar to the one obtained from the 
BFKL equation as shown in
\cite{Martin} for $k_t^2$ as the scale in $\alpha_s$.  
\section{Acknowledgments}
I am very grateful to B. Webber for providing me with
the code of SMALLX. 
I am grateful to
B. Andersson,
 G. Gustafson, 
 H. Kharraziha, J. Kwiecinski, 
L. L\"onnblad, A. Martin, 
S. Munier, 
 R. Peschanski 
and G. Salam 
for many very helpful discussions about CCFM.

\end{document}